\pdfoutput=1
\documentclass{JINST}

\title{Analysis of a Large Sample of Neutrino-Induced Muons with the ArgoNeuT Detector}

\author{C. Anderson$^a$, M. Antonello$^b$, B. Baller$^c$, T. Bolton$^d$, C. Bromberg$^e$, F. Cavanna$^a$$^,$$^f$,
~~~E. Church$^a$, D. Edmunds$^e$, A. Ereditato$^g$, S. Farooq$^d$, B. Fleming$^a$, H. Greenlee$^c$,
~~~~~R. Guenette$^a$, S. Haug$^g$, G. Horton-Smith$^d$, C. James$^c$, E. Klein$^a$, K. Lang$^h$,
~~~~~~~~~~~~~~~P. Laurens$^e$, S. Linden$^a$, D. McKee$^d$, R. Mehdiyev$^h$, B. Page$^e$, O. Palamara$^a$$^,$$^b$$^,$\thanks{Corresponding
author.}, ~~~~~~~~~~K. Partyka$^a$,
G. Rameika$^c$, B. Rebel$^c$, B. Rossi$^g$, M. Soderberg$^c$$^,$$^i$$^,$\thanks{Corresponding
author.}, J. Spitz$^a$, ~~~~~~~~~~~A.M. Szelc$^a$, M. Weber$^g$, T. Yang$^c$, G.P. Zeller$^c$\\
\llap{$^a$}Yale University,
  New Haven, CT  06520  USA\\
\llap{$^b$}INFN - Laboratori Nazionali del Gran Sasso,
  Assergi, Italy\\
\llap{$^c$}Fermi National Accelerator Laboratory,
  Batavia, IL  60510 USA\\
\llap{$^d$}Kansas State University,
  Manhattan, KS  66506 USA\\
  \llap{$^e$}Michigan State University,
  East Lansing, MI  48824 USA\\
    \llap{$^f$}Universita dell'Aquila e INFN,
  L'Aquila, Italy\\
      \llap{$^g$}University of Bern,
  Bern, Switzerland\\
      \llap{$^h$}The University of Texas at Austin,
  Austin, TX  78712  USA\\
   \llap{$^i$}Syracuse University,
  Syracuse, NY  13244  USA\\

  E-mail:\email{ornella.palamara@lngs.infn.it}, \email{msoderbe@syr.edu}}

\abstract{ArgoNeuT, or Argon Neutrino Test, is a 170 liter liquid argon time projection chamber designed to collect neutrino interactions from the NuMI beam at Fermi National Accelerator Laboratory.  ArgoNeuT operated in the NuMI low-energy beam line directly upstream of the MINOS Near Detector from September 2009 to February 2010, during which thousands of neutrino and anti-neutrino events were collected.  
The MINOS Near Detector was used to measure muons downstream of ArgoNeuT.
Though ArgoNeuT is primarily an R$\&$D project, the data collected provide a unique opportunity to measure neutrino cross sections in the 0.1-10 GeV energy range. 
Fully reconstructing the muon from these interactions is imperative for these measurements.  This paper focuses  on the complete kinematic reconstruction of neutrino-induced through-going muons  tracks.    
Analysis of this high statistics sample of minimum ionizing tracks demonstrates the reliability of the geometric and calorimetric reconstruction in the ArgoNeuT detector.}


\begin{document}

\section{Introduction\label{Introduction}}

Liquid Argon Time Projection Chambers (LArTPCs) are well suited for the study of neutrino interactions due to to their unique combination of scalability, fine-grained tracking, and calorimetry.  The LArTPC technique was first proposed in the 1970s \cite{Willis1974221},\cite{rubbia} and has a long history of development~\cite{Chen1978585},\cite{Amerio2004329}.  There is considerable worldwide interest in this technology, with the goal of deploying a multi-kiloton LArTPC as part of a long-baseline neutrino oscillation experiment and proton decay search.   As one step in U.S. based efforts to develop LArTPCs, the ArgoNeuT detector was installed in the MINOS hall,  directly upstream of the MINOS Near Detector (referred to hereafter as the MINOS-ND) \cite{2008NIMPA.596..190T}, at Fermi National Accelerator Laboratory (Fermilab) in the spring of 2009.  The physics run began in September 2009 and lasted about six months, until late February 2010.  ArgoNeuT's data sample is the first collected using the LArTPC technique in the low-energy range relevant for this long-baseline program (0.1-10 GeV).  In fact, only two other LArTPCs have been exposed to neutrino beams, though those detectors were exposed to substantially higher energy beams in the 17-24 GeV range ~\cite{1748-0221-6-07-P07011},\cite{PhysRevD.74.112001}.

In a LArTPC, charged particles traversing a volume of highly purified liquid argon leave a trail of ionization electrons that drift along electric field lines towards a set of instrumented readout planes.  The readout planes consist of finely spaced (mm-scale) wires, with neighboring planes oriented at varying angles to provide independent views of each interaction.  Each wire is read out by a low-noise amplifier sampled in time (MHz-scale) by a digitization board. Combining the views of the ionization electron tracks from these wire planes, by utilizing the common timing information they share, provides a three dimensional reconstruction of the charged particles from the neutrino interaction as well as calorimetric information.

Simulating and reconstructing interactions in these detectors is challenging since the fine-grained tracking and calorimetric aspects of LArTPCs provide substantial information on each neutrino event.  Taking full advantage of this information requires a precise, efficient, and automated simulation and reconstruction package.   The LArTPC technique in ArgoNeuT is augmented by the presence of the MINOS-ND, located just downstream, which is an essential component in the experimental layout and is used to determine the momentum and sign of escaping muons in the neutrino event reconstruction.

In this paper we describe how muon tracks produced by neutrino interactions outside of the ArgoNeuT detector are fully reconstructed and analyzed.  An efficient full kinematic reconstruction of muon tracks exploiting the ArgoNeuT and MINOS-ND capabilities is mandatory for subsequent neutrino cross section measurements, so the work described in this paper lays the groundwork for such future measurements. 

Sections \ref{Detector} and \ref{datasample} describe the ArgoNeuT detector and the data sample under consideration. Sections \ref{recon} and \ref{calor} give an overview of the analysis tools developed for reconstructing particle tracks in ArgoNeuT.   Finally, section \ref{MINOS} describes the full reconstruction and sign-selection of muons using the combination of ArgoNeuT and the MINOS-ND.

\section{The ArgoNeuT detector\label{Detector}}

\begin{figure}[tb]
\begin{center}
\includegraphics[width=3.6in]{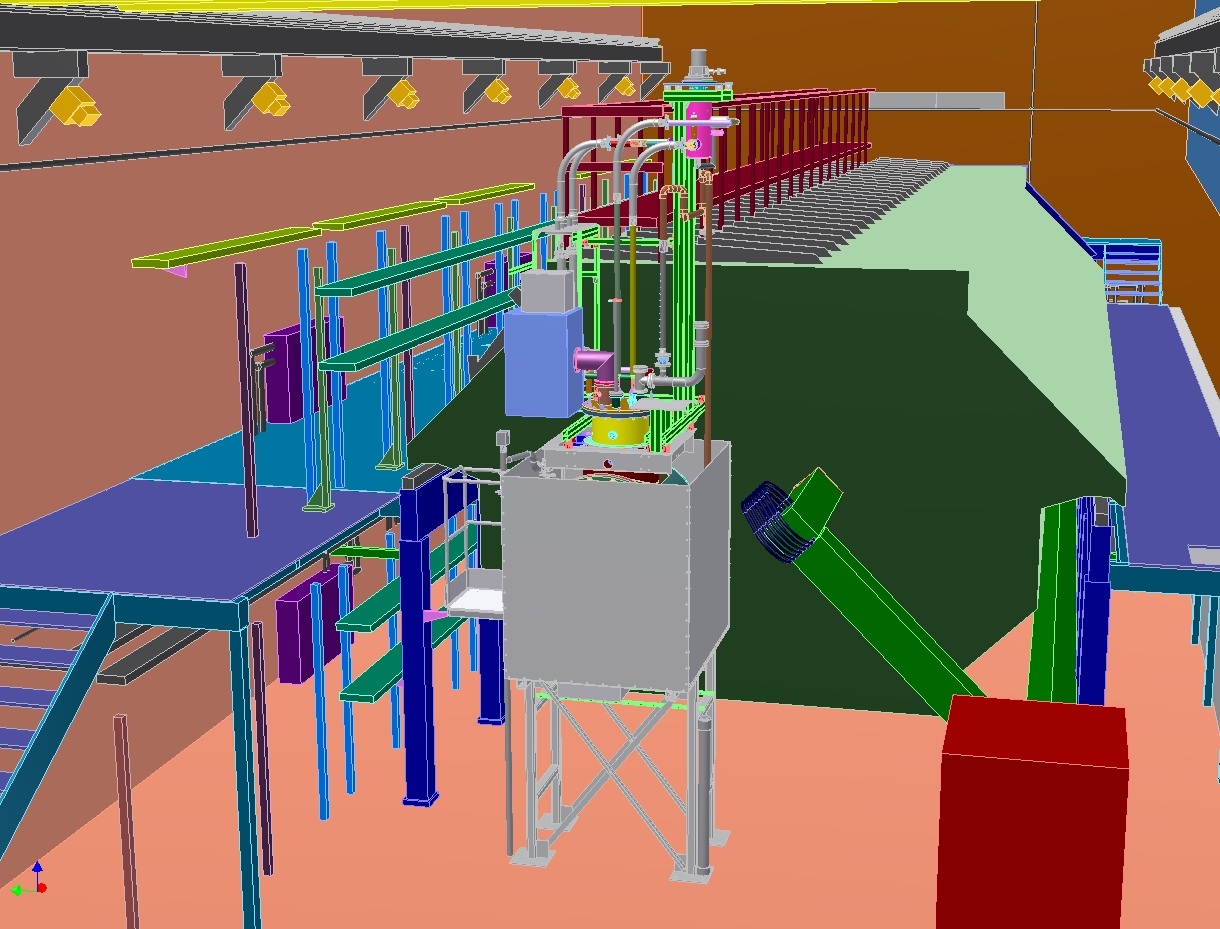} 
\includegraphics[width=2.2in]{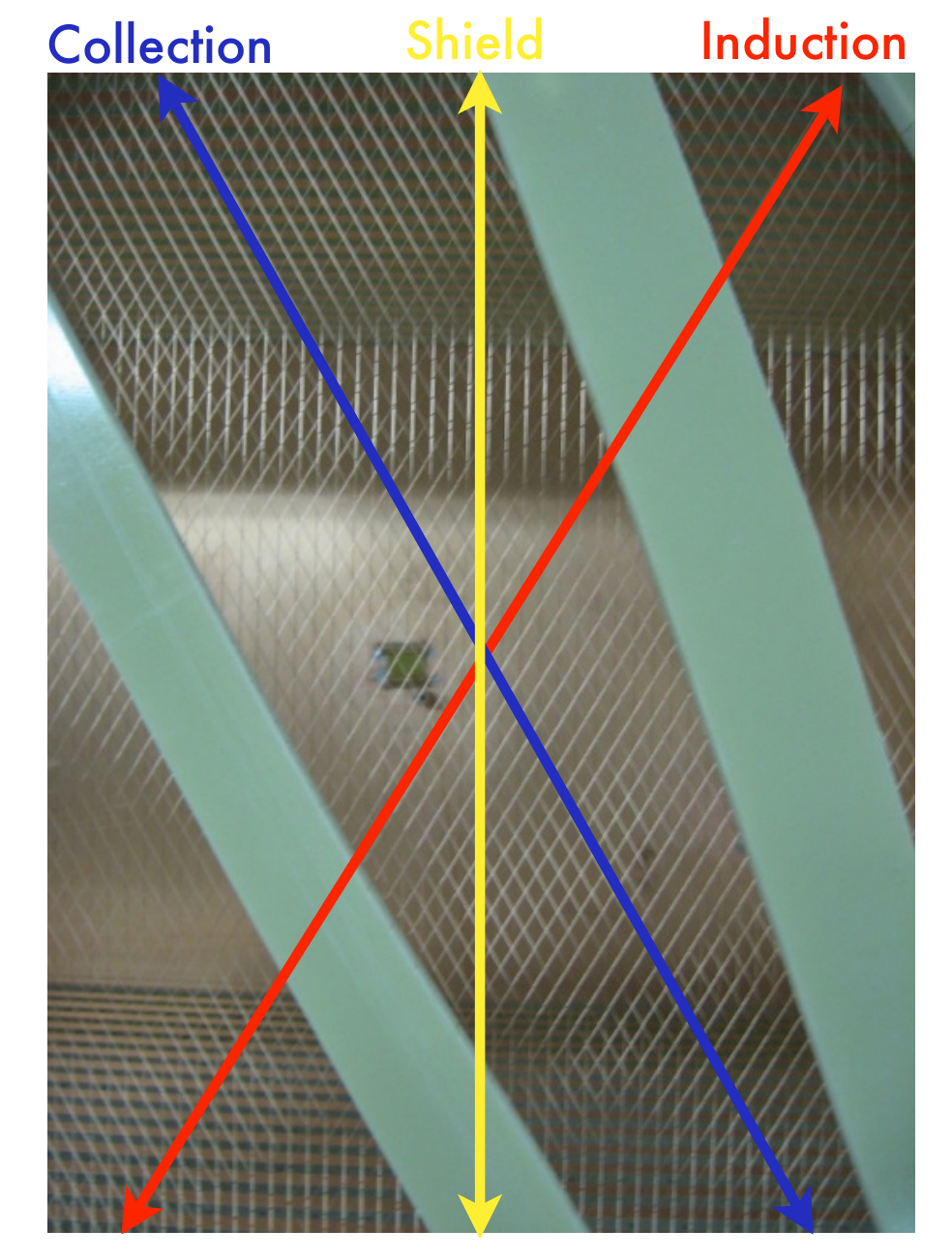} 
\caption{Left: A rendering of the MINOS-ND hall. ArgoNeuT, inside the gray box, can be seen just upstream of the MINOS-ND. ArgoNeuT's TPC position corresponds approximately to the center of the NuMI beam.  Right: Looking through the ArgoNeuT TPC wireplanes, showing the orientations of the wires.}
\label{minos_hall}
\end{center}
\end{figure}

The ArgoNeuT detector features a 550 liter vacuum-insulated cryostat that contains a rectangular TPC box enclosing 170 liters of liquid argon.  The TPC has a maximum drift length of 47.5 cm from the cathode to the first anode plane, and is operated at an electric field of 481 V/cm.  There are three anode planes in the TPC, shown on the right in figure \ref{minos_hall}.  The innermost plane has 226 vertically oriented wires that are not instrumented for electronic readout.  This plane serves to shield the outer wire planes from the drifting ionization in the TPC volume, better localizing their signals in time.  The middle induction plane consists of 240 wires oriented at +60$^{\circ}$ from the beam direction, while the outer collection plane consists of 240 wires oriented at -60$^{\circ}$ from the beam direction. The wire separation, or pitch, in all planes is 4 mm and the planes are spaced 4 mm apart. 

The analog readout electronics consist of a dual JFET front-end integrating preamplifier on a 16-channel card, followed by high-pass and low-pass filters.  These components are located in a double-walled RF-shield Faraday cage. Wire signals are digitized at $\simeq$5~MHz for 406 $\mu$s, starting 10 $\mu$s before the $\simeq$10~$\mu$s-window neutrino beam spill.  The maximum drift time of ionization electrons is approximately 302 $\mu$s between the cathode and the collection plane at the normal operating electric-field value. The wire planes are biased at -298~V (shield), -18~V (induction), and +338~V (collection) with the voltages delivered via 100 M$\Omega$ resistors on each wire.  The bias voltages are chosen to provide constant drift-velocity up to the innermost plane, and maximal transparency for the ionization drifting between the subsequent planes. 

ArgoNeuT uses a closed-loop recirculation system to continually purify and maintain a clean and constant volume of liquid argon.  A 330~W Gifford-McMahon cryocooler (at liquid argon temperatures) mounted several meters above the cryostat is used to re-condense boil-off argon vapor from the liquid volume.  Purification is achieved by directing the re-condensed liquid exiting the cryocooler into one of two pathways, each containing a filter that removes electronegative impurities~\cite{Curioni2009306}.  The purified liquid emerging from the filters is directed back into the cryostat and enters the bottom of the main liquid volume after passing through a sintered metal cap.

 The ArgoNeuT detector was located just upstream of the MINOS-ND in the NuMI neutrino beam \cite{2008NIMPA.596..190T}. ArgoNeuT's location in the MINOS-ND hall is shown on the left in figure~\ref{minos_hall}. 
 Since ArgoNeuT's TPC is too small to contain the majority of the muons produced in neutrino
interactions from the NuMI beam ($\langle E_{\nu_\mu} \rangle\simeq$ 4 GeV in neutrino mode), information from the MINOS-ND is used in the
ArgoNeuT data analysis. This is an enormous advantage for ArgoNeuT since the MINOS-ND, in addition to providing the momentum of the escaping muon, can also determine the sign of the muon with its magnetized detector.

A detailed technical description of the ArgoNeuT detector, the commissioning, the data taking and the off-line event reconstruction can be found in Reference \cite{techpaper}.  The data analysis is also briefly described in section~\ref{recon}.

%

\section{Neutrino-induced through-going muon data sample\label{datasample}}

During ArgoNeuT's physics run about $1.35\times10^{20}$ protons on target (POT) 
were delivered to the NuMI target which was operating in the "low-energy" configuration \cite{Anderson:1998zz}. 
The data analyzed in this paper comes from an exposure taken in September 2009, while the NuMI beam was in neutrino-mode, during which 8.5$\times$10$^{18}$ POT was acquired.  Though neutrino-mode data is analyzed in this paper, both neutrinos and antineutrinos are present in the sample due to the small contamination of antineutrinos present in the beam (92.9\% $\nu_{\mu}$, 5.8\% $\overline{\nu}_{\mu}$, 1.3$\%$ $\nu_e + \overline{\nu}_e$) \cite{PhysRevD.81.072002}.

Every spill from the NuMI beam was recorded, using the accelerator clock signal as a trigger to initiate readout.  The sample collected includes (i) neutrino events in which an  interaction vertex with one or more tracks is associated, and (ii) "through-going track(s)" events where charged particles produced by a neutrino interaction upstream of the ArgoNeuT detector propagate through the LArTPC volume.  The through-going track events comprise the majority of the data sample and are mainly muon tracks, since other particles are stopped before reaching the ArgoNeuT detector. In particular, due to the large amount of material in front of the ArgoNeuT detector most pions created upstream do not reach ArgoNeuT.  As a result, the $\pi$/$\mu$ misidentification rate is small for the present study, even when accounting for possible pion decays in flight in the liquid argon volume.

\begin{figure}[tb]
\begin{center}
\includegraphics[width=6.0in]{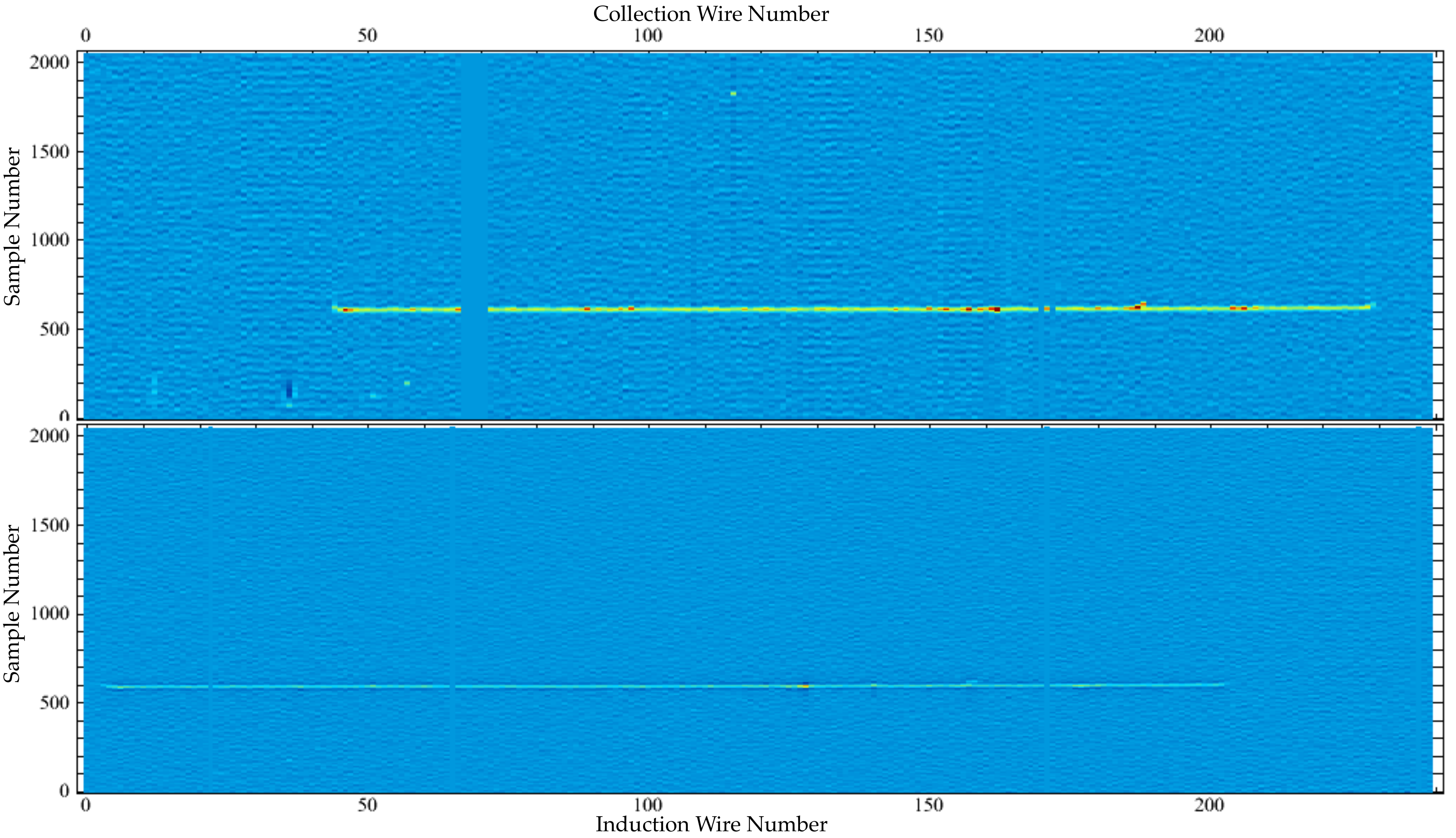}
\caption{A neutrino-induced through-going track as seen in ArgoNeuT's induction (bottom) and collection (top) plane views.  The horizontal axis of these displays corresponds to wire number within a wireplane.  The vertical axis corresponds to the sampling time elapsed after the start of the trigger, which can be converted into the drift distance where the ionization originated.  The beam enters from the left-hand side of this image, and the MINOS-ND is located on the right-hand side.  The fully reconstructed version of this track, including MINOS-ND information for this event, is shown in figure 3.} 
\label{ThroughGoingMu}
\end{center}
\end{figure}

\begin{figure}[!h]
\begin{center}
\includegraphics[width=5.2in]{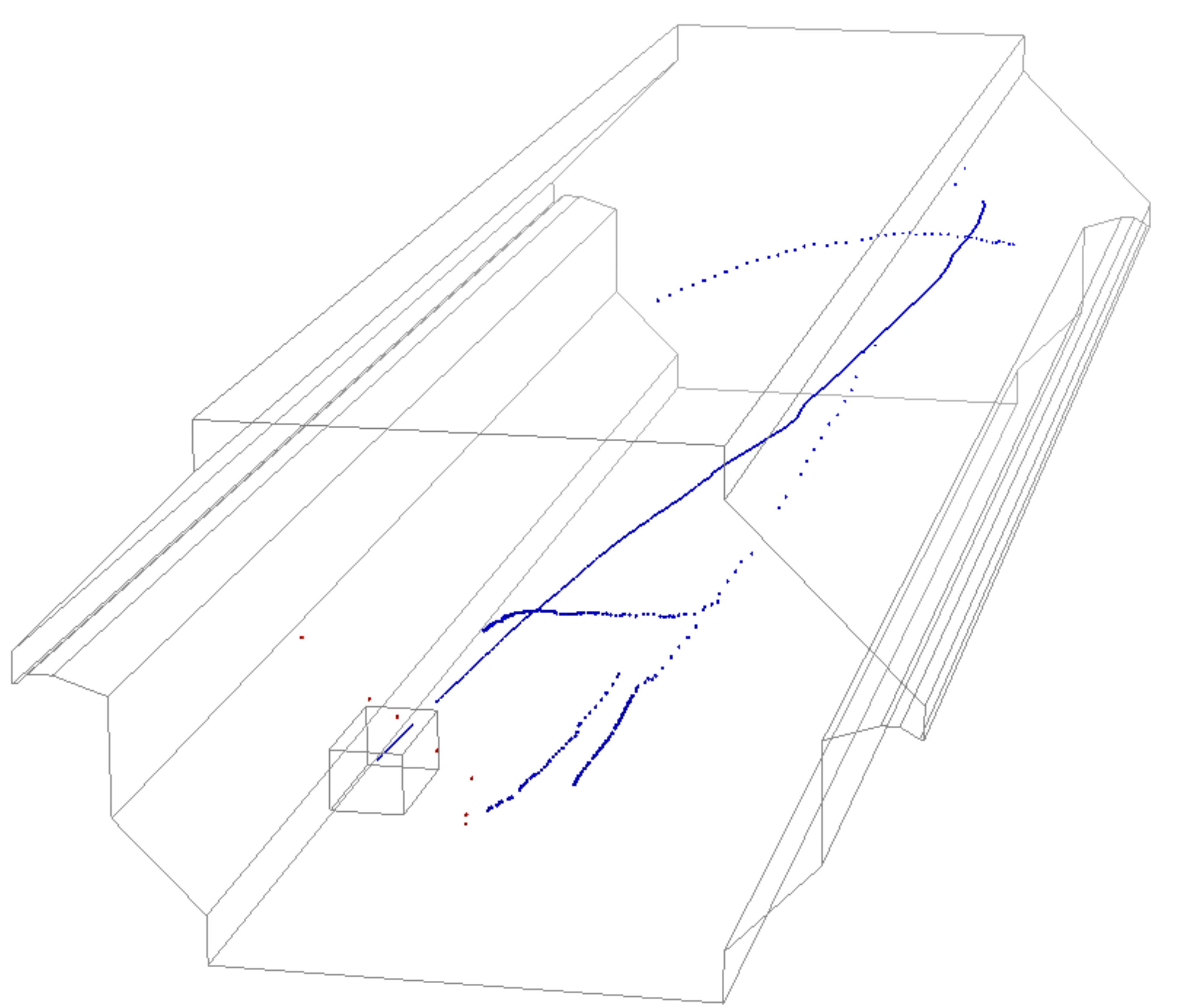}
\caption{Display of the ArgoNeuT and MINOS-ND geometries using alignment from empirical measurement.  In the image shown, a through-going track in ArgoNeuT has been matched to a negatively charged track in the MINOS-ND.  Other tracks present in the MINOS-ND during the spill are also represented.}
\label{coordinates}
\end{center}
\end{figure}

Events with through-going tracks created by neutrino interactions occurring upstream of ArgoNeuT have been selected through a fully automatic procedure, using the topology and timing of ionization depositions within the TPC volume.  Figure \ref{ThroughGoingMu}  depicts an example of the raw data for a through-going track as seen by ArgoNeuT's two wire planes, while figure \ref{coordinates} shows the same track fully reconstructed along with the corresponding information from the MINOS-ND.

The selected through-going events have been automatically reconstructed, as described in section \ref{recon}.  The ArgoNeuT analysis tools are part of a general software toolkit called LArSoft \cite{larsoft} used to reconstruct particle tracks in LArTPCs.  LArSoft has been developed in collaboration with MicroBooNE \cite{microboone} and LBNE \cite{Akiri:2011dv}. 

\section{Geometric reconstruction of through-going tracks\label{recon}}

Each muon in the through-going data sample crosses a large number of wires in the detector and can be reconstructed as a line-like track.  The complete reconstruction procedure has been applied to the neutrino-induced through-going muon data sample. The raw data from the detector is first calibrated to remove any baseline shifts and also to convert the induction plane pulses from bipolar (resulting from the induction signal from the passing ionization electrons) to unipolar shapes.  The calibrated signals from all wires are then scanned for hits, or localized depositions of charge above threshold, which are then parameterized by a Gaussian fit.  Proximal hits in both views are grouped together into clusters, which are further fit to straight-line trajectories using a Hough transform \cite{hough}.  

Three dimensional (3D) tracks are reconstructed by combining associated two dimensional (2D) line-like clusters that are identified in both views of the TPC.  Geometric parameters of the track such as direction cosines and track pitch length are reconstructed, where track pitch length is defined as the effective length of the portion of track visible to a single wire.  Once a 3D track is identified, a hit-by-hit association procedure is applied to match hits from the two wire planes to obtain a fine-grained 3D image of the event.  This last step is based on a dedicated matching algorithm to form 3D points, that can also be applied to the 3D reconstruction of arbitrary trajectory tracks that are not necessarily straight lines. 

\begin{figure}[h]
\begin{center}
\includegraphics[width=5.0in]{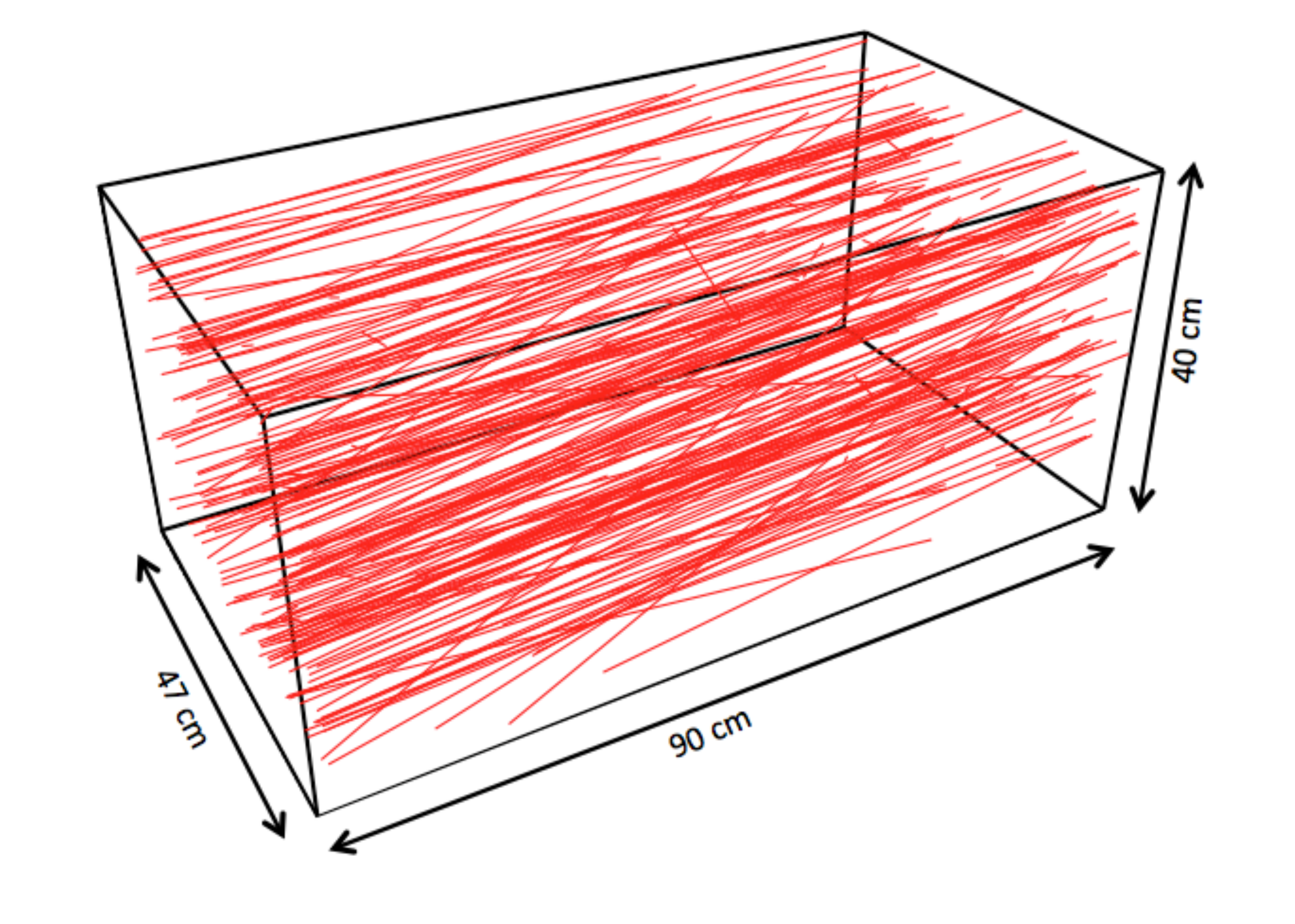}
\caption{ 3D reconstructed tracks for a sample of overlaid through-going muon tracks crossing the ArgoNeuT active volume.}
\label{event3D}
\end{center}
\end{figure}

Tracks having both the first and the last reconstructed points falling within 1.5 cm of a TPC boundary, which are also matched to a track reconstructed by the MINOS-ND (see section \ref{MINOS}) are included in the through-going data sample.  Figure~\ref{event3D}  shows many examples of the reconstruction of a few hundred overlaid through-going muons.  The vast majority ($>$95\%) of these tracks are $\sim$90 cm in length, consistent with entering through the upstream face of the TPC and exiting through the downstream face.  The rest of the tracks in the sample enter or exit the TPC through one of the side walls, and can be significantly shorter than 90 cm in length.  The final sample contains 14322 negatively-charged through-going tracks, and 2607 positively-charged through-going tracks.   The ratio of $\mu^+/\mu^-$ in the sample (15.4$\%$/84.6$\%$) is higher than would be expected by comparing the breakdown of $\overline{\nu_{\mu}}/\nu_{\mu}$ neutrinos in the NuMI flux for neutrino mode (6$\%$/94$\%$)  \cite{PhysRevD.81.072002}.  This is not unexpected since the through-going sample under consideration has a higher average energy than that of muons from the overall flux, and the ratio of  $\mu^+/\mu^-$ increases with increasing energy.  


\begin{figure}[h]
\begin{center}
\includegraphics[height=2.05in]{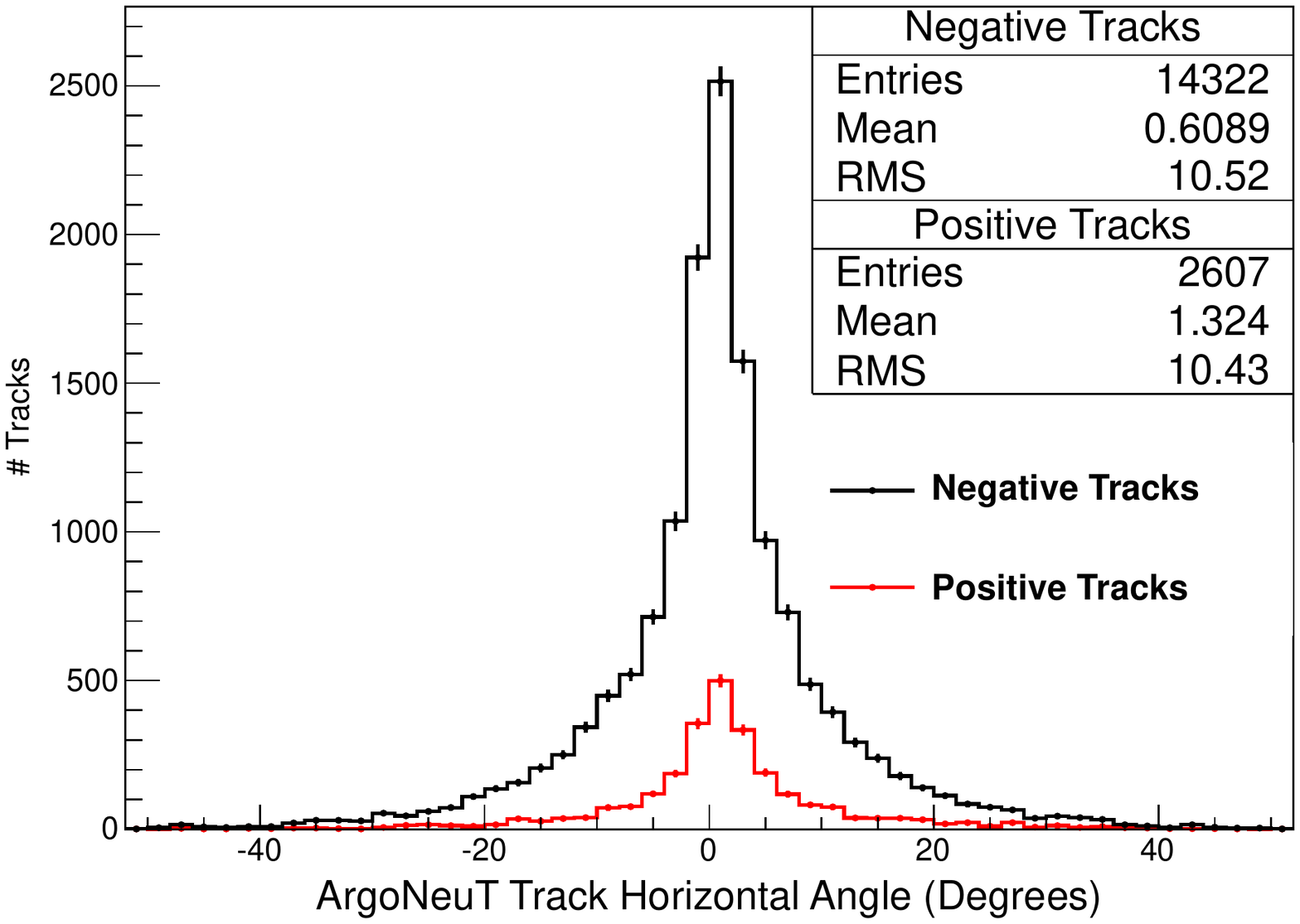} 
\includegraphics[height=2.05in]{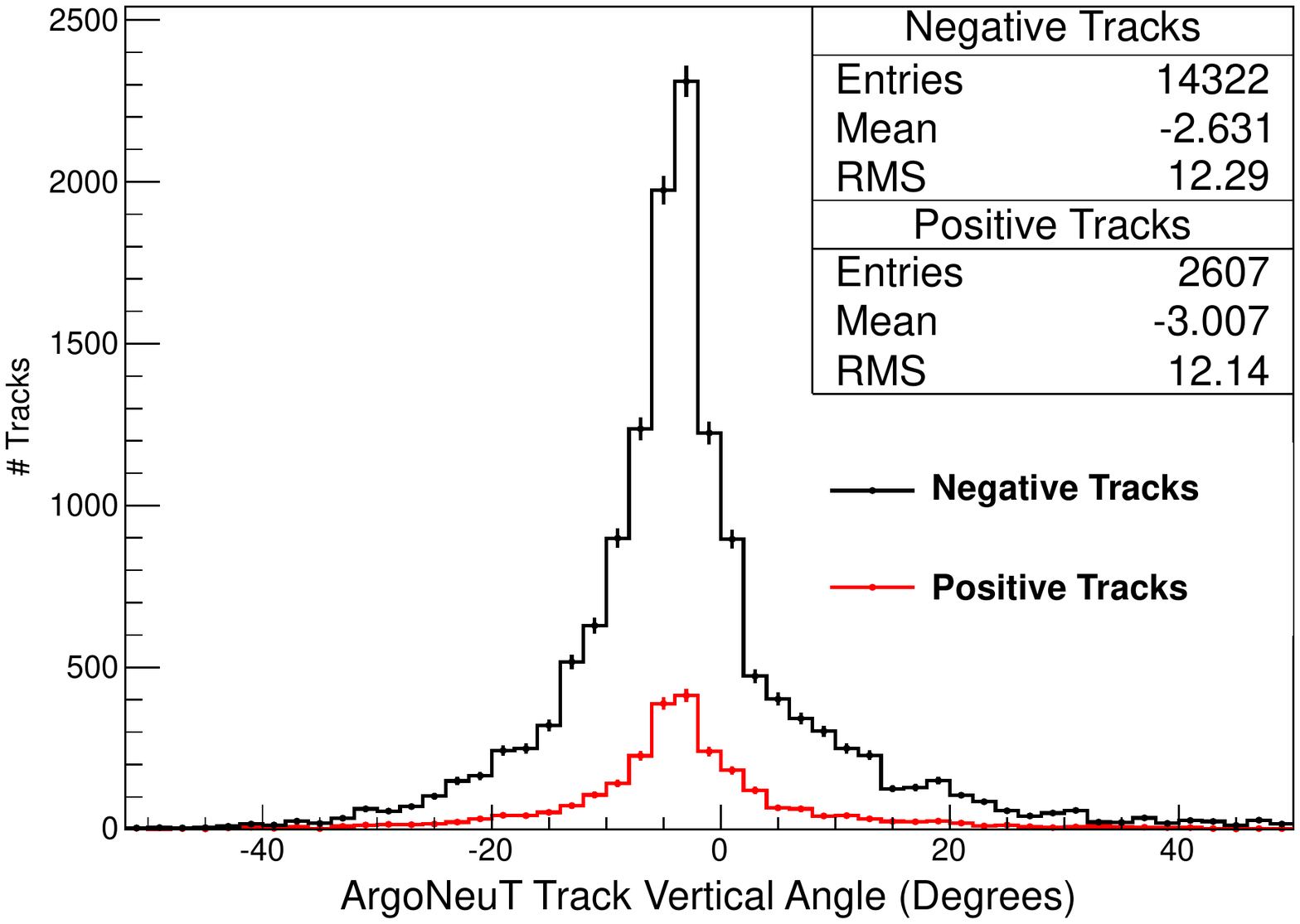} 
\caption{Horizontal (left) and vertical (right) angles of reconstructed tracks in ArgoNeuT, for positively and negatively charged tracks (using sign determined by the MINOS-ND).}
\label{angles}
\end{center}
\end{figure}

In figure~\ref{angles}, the inclination of the reconstructed tracks with respect to the horizontal (left) and vertical (right) axes is shown. The mean value of the horizontal angle (0.61$^\circ$ for negatively-charged tracks) suggests a slight misalignment of the ArgoNeuT TPC with respect to the neutrino beam direction.  The mean vertical angle is peaking at approximately -3$^\circ$, which is consistent with the NuMI beam's known inclination angle with respect to ArgoNeuT and the MINOS-ND as it heads down through the decay tunnel towards Minnesota. 


\section{Calorimetric reconstruction of through-going muons}
\label{calor}
The 3D geometric reconstruction described in section \ref{recon} is a preliminary step for the calorimetric measurement of the energy released in liquid argon.  This calorimetric measurement is performed by first accounting for the charge loss due to electro-negative impurities, and then converting charge to energy after correcting for the quenching effect in which liberated ionization recombines with argon ions.

The hit amplitude (in units of ADC counts) is converted to charge (in units of number of electrons) and then normalized for the track pitch length to obtain the charge deposited per unit length.  For the through-going muon data sample the reconstructed track pitch length in the collection wire plane has an average length of (0.49$\pm$0.07)~cm, reflecting the fact that most of the through-going muons are nearly parallel to the wire planes.  An electronic calibration factor $f_{cal}$ =7.6 ADC/fC (determined at design level and confirmed from test bench measurements of the ArgoNeuT electronics \cite{techpaper}) has been applied to convert from ADC counts to charge 
expressed in number of electrons.


\subsection{Charge to energy conversion}

To account for the charge loss along the drift due to electro-negative impurities the charge deposited per unit track length is multiplied by $e^{\Delta t/\tau}$, where $\Delta t$ is the hit drift time  and $\tau$ is the measured electron lifetime in ArgoNeuT (typically $\tau\sim$700~$\mu$s during the period of running covered here).   

 Muon tracks can be accompanied by the emission of knock-on electrons ($\delta$-rays) along the track. In the calorimetric reconstruction procedure, an algorithm is used to identify and properly take into account the contribution due to $\delta$-rays.
 Low energy $\delta$-rays tend to be emitted parallel to the muon direction, overlapping with the parent muon track, 
 with ionization charge from both being focused onto a single wire hit  (the hit amplitude in this case is approximately twice the muon hit amplitude). 
 Energetic $\delta$-rays span over few subsequent wire hits and the most energetic  ones (a few MeV)  
 tend to move away from the muon track, producing a spatially resolved nearby hit on the same wire.
 Once the hits belonging to the $\delta$-ray are identified, their contribution (charge of the hit - subtracted by the average contribution corresponding to a muon in case of overlapping hits) is assigned to the first hit where the origin of the $\delta$-ray electron is located.  
 
The $dQ/dx$ distribution (in $e^-$/cm)  obtained from the through-going track sample, having taken into account the contribution due to $\delta$-rays (in about 13\% of the hits), is well fit  with the convolution of a Landau function with a Gaussian function.  The Landau function describes the energy loss by the charged particles traveling through a medium, while the Gaussian encompasses the effects of electronic noise, ionization diffusion, and track orientation with respect to the wires, among others.  The $\delta$-ray removal algorithm improves the $\chi^2$/D.O.F. of the Landau+Gaussian fit by a factor two, while the best fit parameters remain unchanged within statistical uncertainty.  The $\sigma$ parameter from the Gaussian fit is 4259 electrons per centimeter.  This corresponds to a contribution of about 1700 electrons to the spread of the single wire signal, mainly coming from electronic noise.  

\begin{figure}[tb]
\begin{center}
\includegraphics[width=6in]{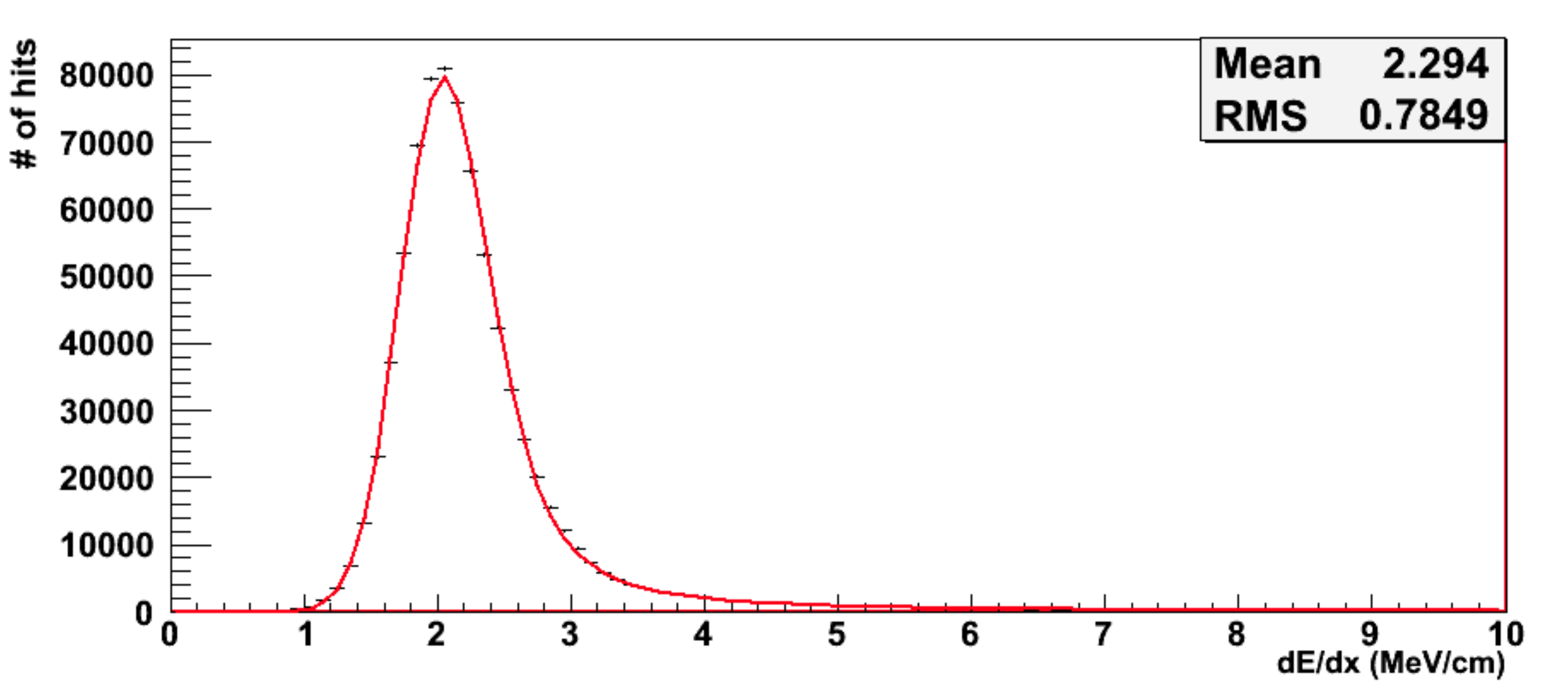}
\caption{ Energy per unit track length deposited by the beam-induced through-going muons in ArgoNeuT, corrected for the contribution of $\delta$-rays.  The error bars shown are statistical only.  The results from a Landau-Gaussian fit (shown in red) are also reported.}
\label{dedx}
\end{center}
\end{figure}

Finally, the full calorimetric reconstruction of the events is performed accounting for the
quenching effect on the ionization charge, using the semi-empirical Birks model~\cite{birks} to convert  $dQ/dx$ into energy released per unit track length $dE/dx$.  
This is done using the parameterization reported in ~\cite{2004NIMPA.523..275A}, where 
the dependence of the recombination on the particle stopping power in liquid argon has been fit with a  Birks functional dependence, as a function of the drift electric field. 


The $dE/dx$ distribution measured for the through-going muon sample is shown in figure~\ref{dedx}. The average energy release is $\langle dE/dx \rangle=$2.3$\pm$0.2 MeV/cm, in good agreement with theoretical expectations 
for a sample of muons with an average energy of 7.0 GeV (see figure~\ref{minos_dist}). The quoted errors include a negligible statistical uncertainty as well as systematic uncertainty due to the errors on the electronic calibration factor, the Birks parameters, the electric field in the drift region, the measurements of the electron lifetime, and the track pitch length.
The $dE/dx$ distribution is very well described by a convolution of a Landau function with a Gaussian function.  The through-going muons are measured to deposit approximately 200 MeV of total energy in the TPC. 

\subsection{Detector self-calibration}
In the previous section, the procedure to measure the most probable value of the energy deposited by the through-going muons has been described, starting from raw data and using the electronic calibration factor of the detector.  In the present section, the inverse procedure, wherein we assume the theoretically most probable value of the deposited energy in order to determine the detector calibration factor, is described.

After correcting for the electron lifetime and quenching effect on the ionization charge and properly taking into account the contribution due to $\delta$-rays, as reported in the previous section, the $dQ_0/dx$ distribution is fit with a convolution of a Landau and a Gaussian function.  The most probable (m.p.) value of the Landau distribution is obtained:
\begin{equation}
\label{equation:dQ0dx}
\frac{dQ_0}{dx} \bigg\vert_{m.p.} =  83\pm 1~\mathrm{ADC/cm}
\end{equation}

\begin{figure}[h]
\begin{center}
\includegraphics[width=5.5in]{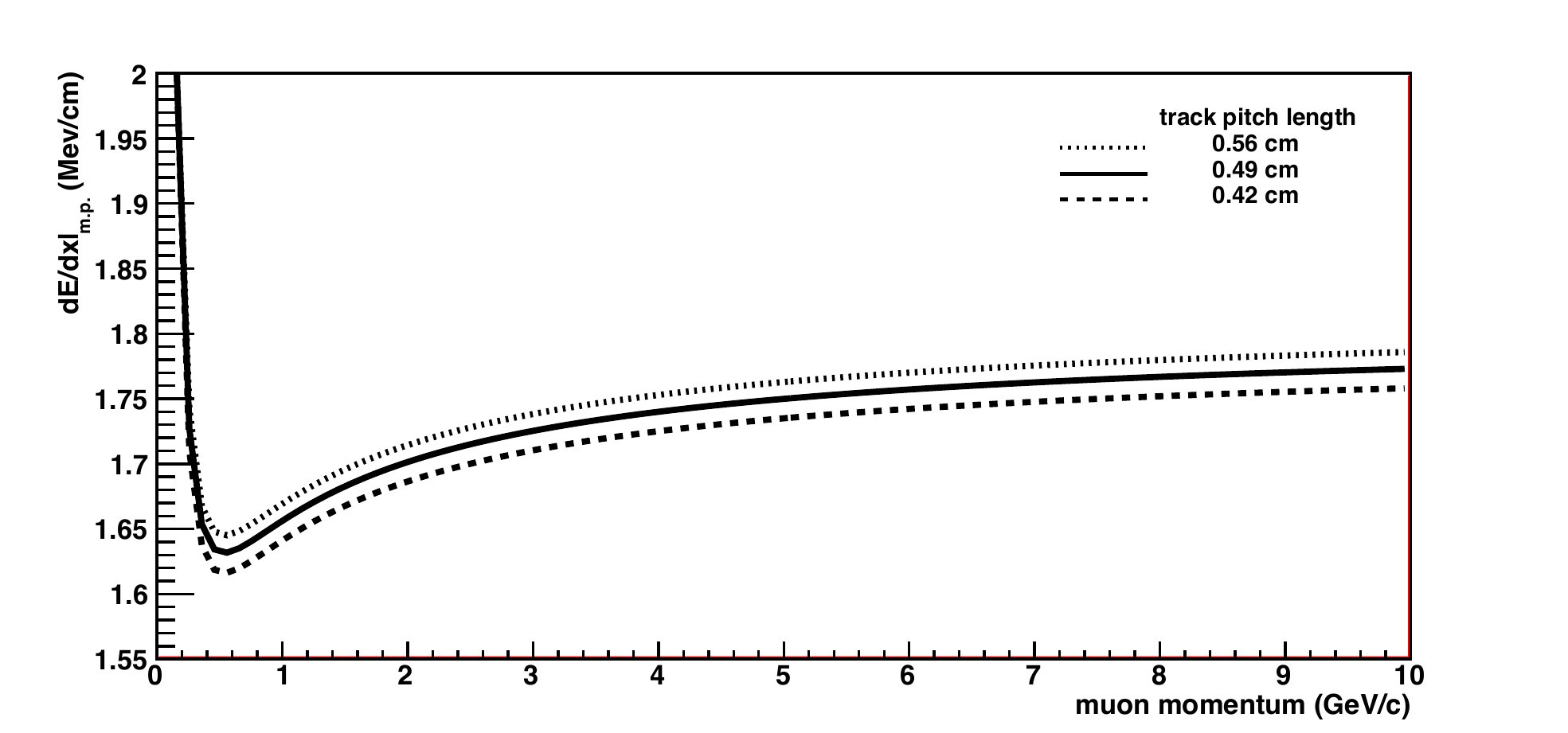}
\caption{Theoretical energy loss as a function of muon momentum for various liquid argon thickness values, corresponding to the range of measured track pitch length, 0.49$\pm$0.07~cm.}
\label{mpv}
\end{center}
\end{figure}

Neutrino-induced through-going muons crossing the detector have an average energy of 7.0~GeV and a momentum spectrum as reported in section~\ref{MINOS}. 
In a dense medium such as liquid argon, the value of the energy loss depends both on the particle momentum and on the absorber thickness. In the momentum range of interest, a MC simulation has been performed to evaluate the theoretical energy loss for muons traversing different values of liquid argon thickness, as shown in figure~\ref{mpv}, that correspond to the average value of track pitch length for this data sample, 0.49$\pm$0.07~cm. 
Convolving the theoretical energy loss, as given in figure \ref{mpv}, with the measured energy spectrum (reported in figure \ref{minos_dist}) provides an estimation of the mean value of the most probable energy loss
for the through-going muon sample:
\begin{equation}
\label{equation:dEdx}
\frac{dE}{dx}\bigg\vert _{m.p.}^{th} = 1.73 \pm 0.02~ \mathrm{MeV/cm}
\end{equation}
From the ratio of equations~\ref{equation:dQ0dx} and \ref{equation:dEdx} an electronic calibration factor 
\begin{equation}
f_{self-cal}=50.8 \pm 0.6 ~\mathrm{ADC/MeV}
\end{equation}
is obtained. By using the ionization potential in liquid argon (1 e$^-$=23.6 eV) we obtain:
\begin{equation}
f_{self-cal}=7.49 \pm 0.09~\mathrm{ADC/fC}
\end{equation}
This value is in good agreement with that obtained from completely independent test bench measurements of the electronics, demonstrating the reliability of the ArgoNeuT calorimetric reconstruction.

\section{Full kinematic reconstruction of through-going muons\label{MINOS}}
The modest size of the ArgoNeuT TPC makes it necessary to use the downstream MINOS-ND to fully reconstruct the kinematic properties of tracks that are detected but not contained by ArgoNeuT, which is the case for all through-going muons.  The magnetized MINOS-ND also provides sign identification for the muons, which is crucial for cross-section measurements \cite{Anderson:2011ce}.  

The MINOS collaboration has provided reconstruction information for tracks in their detector during each NuMI spill recorded while ArgoNeuT was in operation.  This information is associated offline with ArgoNeuT data from the same spill using a beam timestamp common to events in both detectors.  All tracks identified by ArgoNeuT and the MINOS-ND in the same NuMI spill are then searched to identify corresponding tracks in the two detectors.

ArgoNeuT tracks are extrapolated along their directional three-vector from their exit position in the TPC to the first plane of the MINOS-ND.  Then, matching criteria is applied which compares the angle between the ArgoNeuT and MINOS-ND tracks as well as the radial difference between the projected-to-MINOS ArgoNeuT track and the candidate MINOS-ND track.  Any MINOS-ND track originating within 20.0 cm (along the beam direction) of this plane, within a transverse radial distance ($\Delta$R) of 35.0 cm of the extrapolated ArgoNeuT track, is ``matched" to the ArgoNeuT track.  The relative alignment  between the two detectors' coordinate systems, which was measured only roughly during the data-taking period, has been found empirically by comparing many ArgoNeuT and MINOS-ND track pairs and optimizing the coordinate alignment to minimize the separation of matched tracks while maximizing the overall number of tracks matched.  

\begin{figure}[!h]
\begin{center}
\includegraphics[width=4.0in]{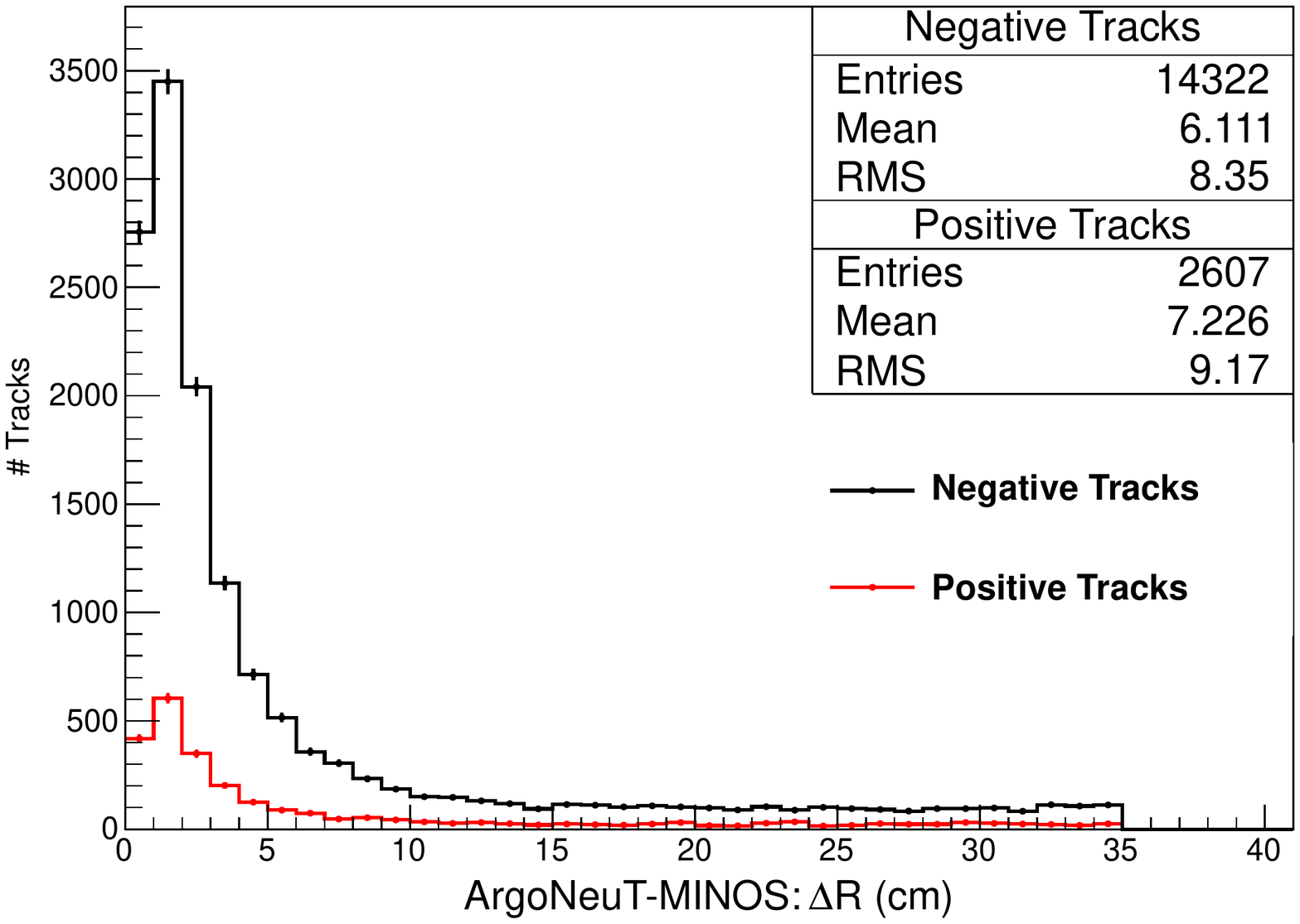} 
\caption{Radial distance between the extrapolated track from ArgoNeuT and the matched track in the MINOS-ND, at the first plane of the MINOS-ND.}
\label{deltaR}
\end{center}
\end{figure}

\begin{figure}[!h]
\begin{center}
\includegraphics[width=1.95in]{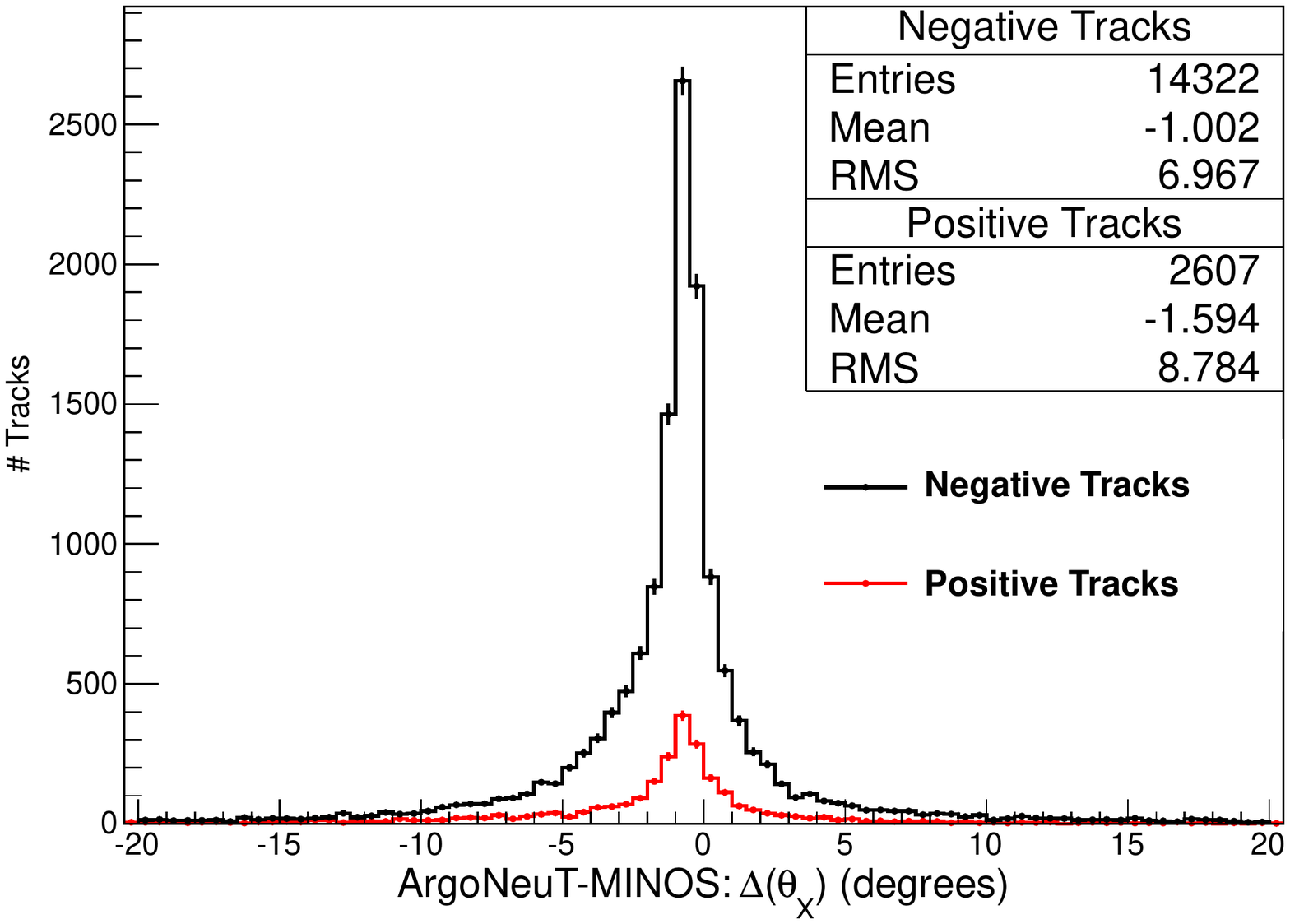} 
\includegraphics[width=1.95in]{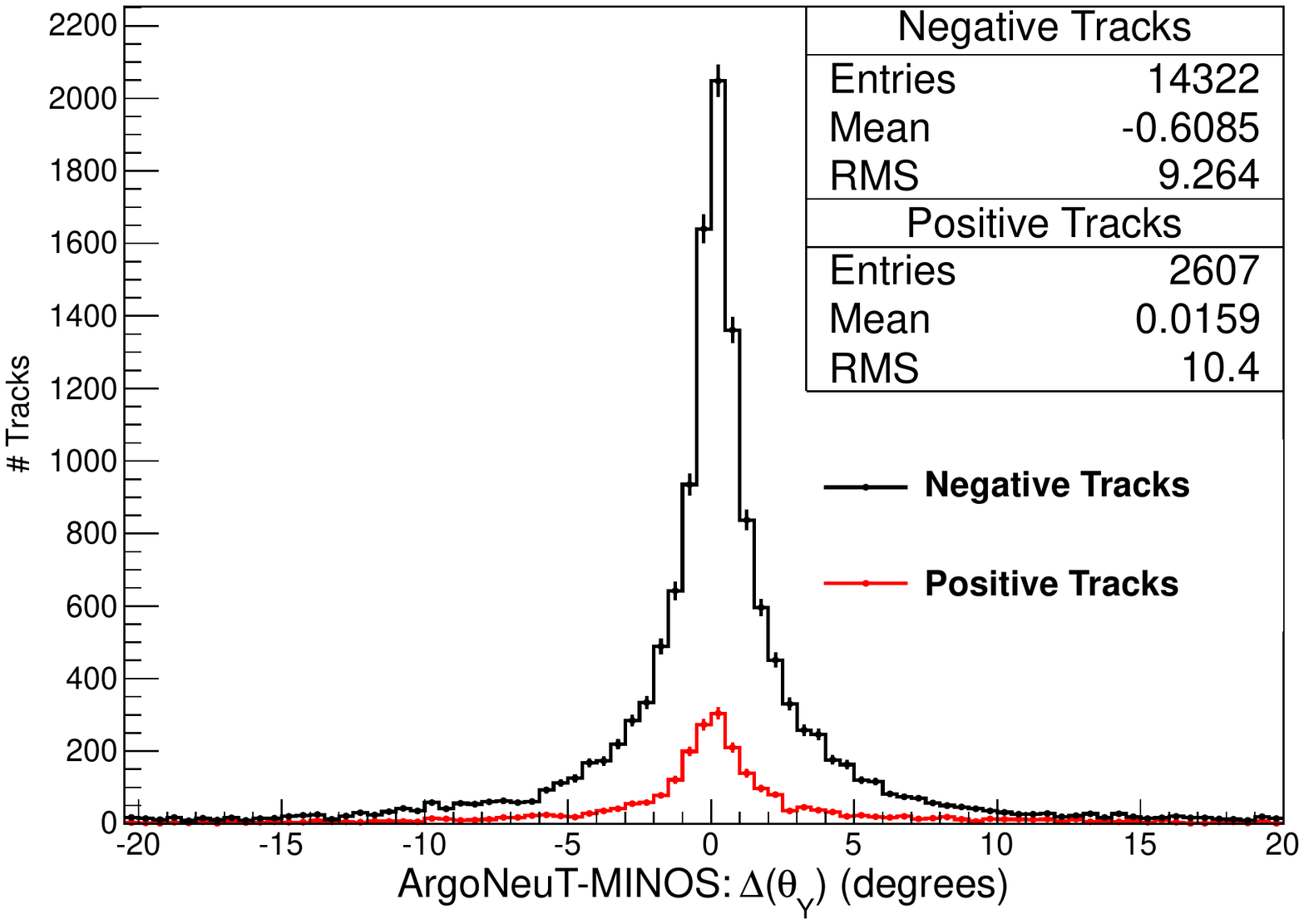} 
\includegraphics[width=1.95in]{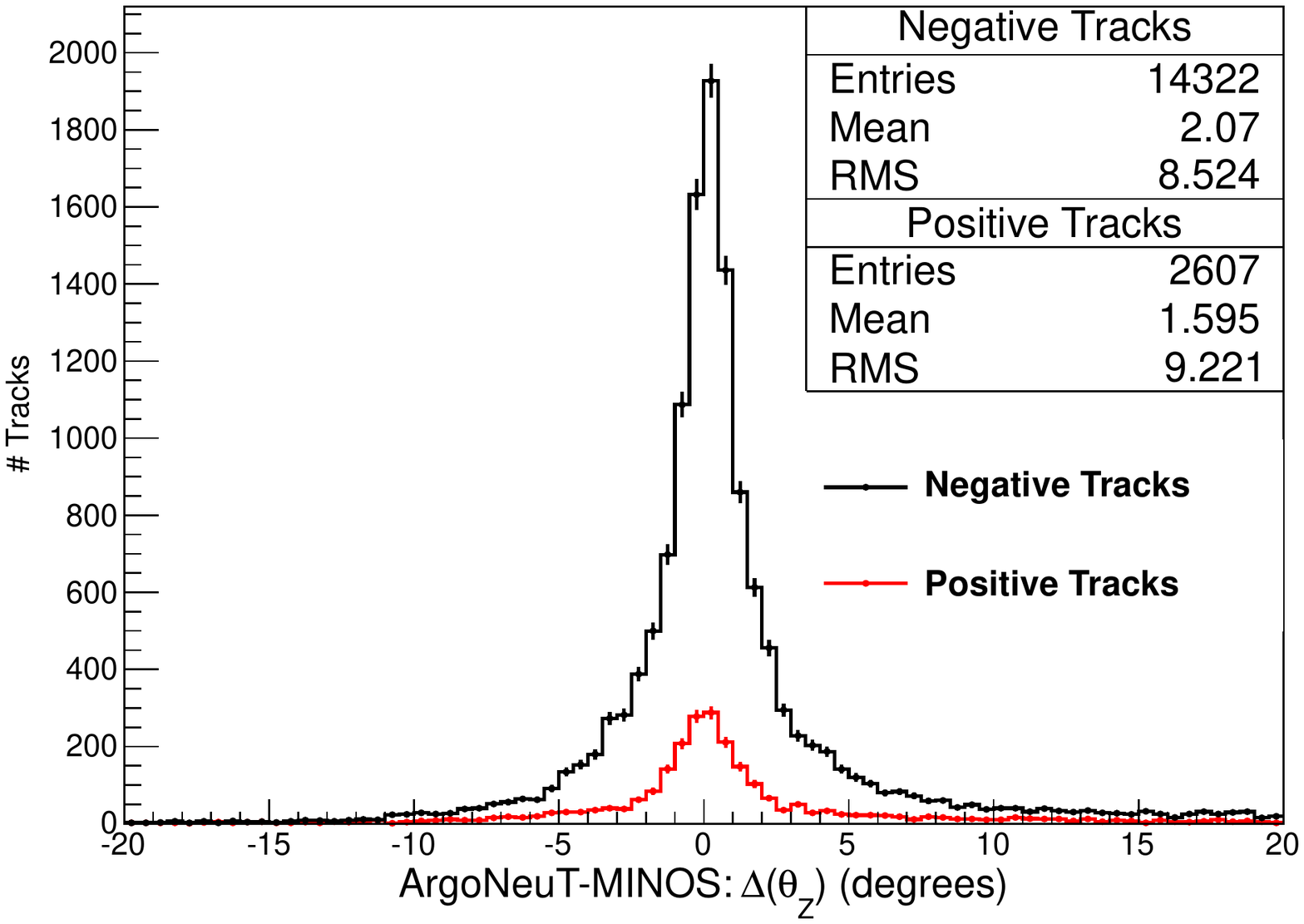} 
\caption{Difference in track direction, as measured by ArgoNeuT and the MINOS-ND for tracks passing the matching algorithm.}
\label{minos_cos_dist}
\end{center}
\end{figure}

This matching routine is applied to all reconstructed ArgoNeuT tracks in the data sample considered.  Figure~\ref{deltaR} shows the $\Delta$R distribution for the projected ArgoNeuT track and the matched MINOS-ND track for all $\approx$17,000 tracks in the sample. The spread in this distribution is primarily due to multiple scattering that occurs after the muon exits the ArgoNeuT active volume but prior to entering the MINOS-ND.  Figure \ref{minos_cos_dist} shows the measured difference in direction between the matched tracks reconstructed in ArgoNeuT and the MINOS-ND.

\begin{figure}[!h]
\begin{center}
\includegraphics[width=4.0in]{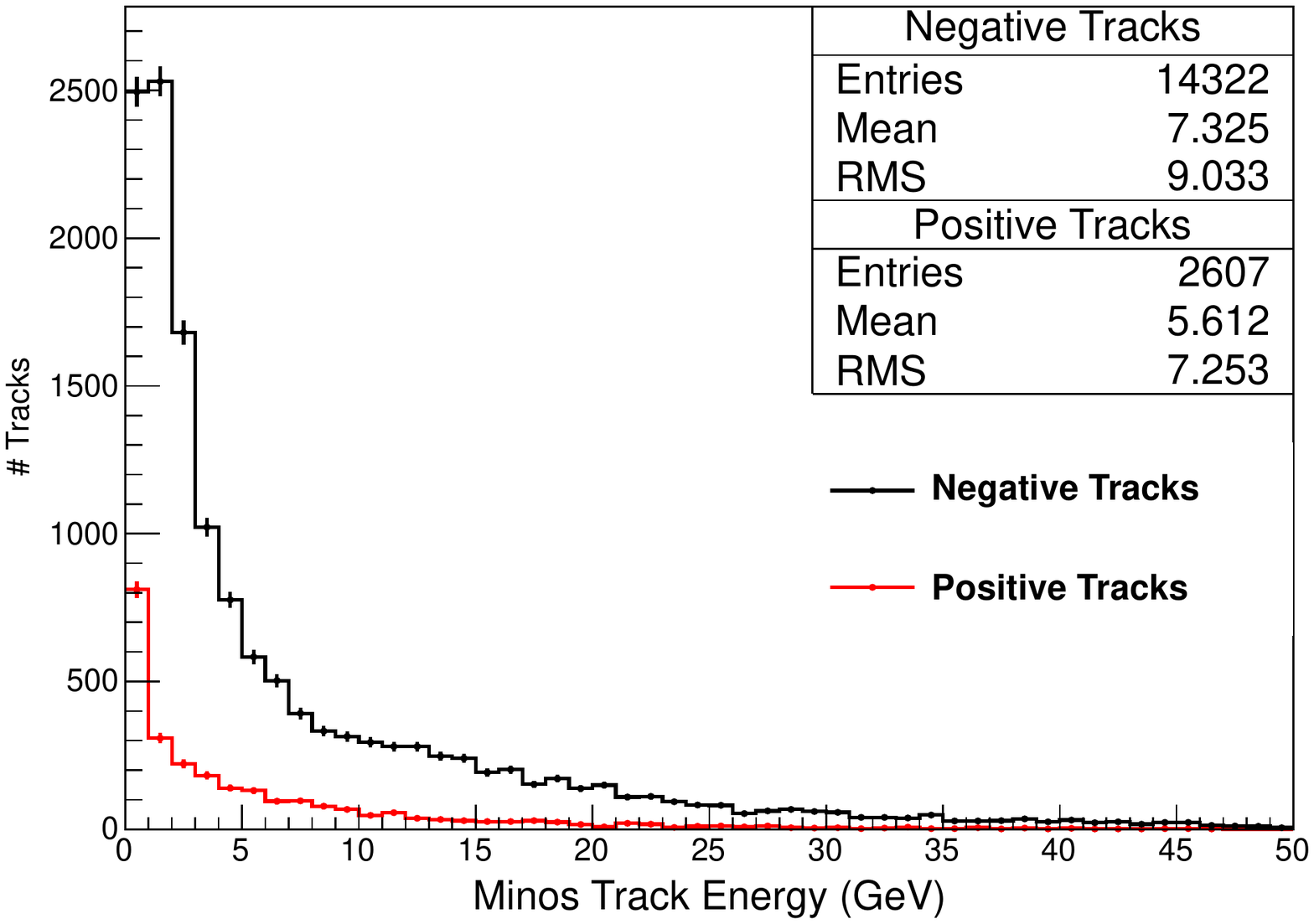} 
\caption{Energy of MINOS-ND tracks that have been matched to ArgoNeuT tracks.}
\label{minos_dist}
\end{center}
\end{figure}

The track momentum is measured from range if the track stops within the MINOS-ND, or from curvature in the toroidal magnetic field if it exits.  Figure \ref{minos_dist} shows the energy measured by the MINOS-ND for tracks matched to through-going ArgoNeuT tracks, separated by the sign of the track.  The energy distribution in figure \ref{minos_dist} is entirely from the MINOS-ND, and does not include the contribution from energy deposited in ArgoNeuT.  The contribution from energy deposited in ArgoNeuT is only $\sim$200 MeV on average, as discussed in section \ref{calor}, which is negligible for most of the muons.

\section{Conclusion\label{conclusion}}

This analysis of through-going track data in ArgoNeuT demonstrates the 3D and calorimetric reconstruction capability of the LArTPC technology, and shows that a precise reconstruction of the event kinematics is achieved.  These techniques form the basis of subsequent cross-section measurements that ArgoNeuT will perform with its entire NuMI data sample \cite{Anderson:2011ce},\cite{spitz}.  The work presented in this paper is an important step in the development of fully automated reconstruction of neutrino interactions for all future LArTPC experiments.  

\newpage

\acknowledgments
We gratefully acknowledge the cooperation of the MINOS collaboration in providing their data for use in this analysis.  We also wish to acknowledge the support of Fermilab, the Department of Energy, and the National Science Foundation in ArgoNeuT's construction, operation, and data analysis.
     
\bibliographystyle{JHEP}
\bibliography{ArgoNeuT_Ref}{}

%
%
%
%
\end{document}